# EDUCATION AND TECHNOLOGY IN KUWAIT DEMOCRATIC PRACTICE

**Fatmah A. Husain**
**Hasan A. Abbas**
**Kuwait University**

*ABSTRACT*

*This study discusses the effects of the new technology on the democratic system by reviewing the political history of democracy in philosophic view. The study explores the philosophical discussion of democracy and its components that is designed by most influential philosophers whom most of them agreed on the importance of education and knowledge to be the two main components. The study connects the democratic progress in the State of Kuwait to its roots, namely knowledge and education.*

**INTRODUCTION**

The information technology was developed remarkably in recent years. Like any other prior manmade technology, information technology was developed to assist humanity seeking for welfare and comfortable life by managing his related matters and in different aspects: temporal, local, political, administrative, scientific...etc. This technology added diversity to our choices and facilitated the manners regarding how we trade our different life requirements. Moreover, this technology provided new ways for living that were undiscovered in previous societies due to their limited abilities at that time.

The political practice was one of the oldest life necessities for the human society in many nations, and for this reason many governments and political regimes were formed in the history. The ruling regimes are named differently according to the governed practice such as the democratic system (one of the most acceptable and widely used political system) with its different types. Another political systems are socialism system, dictatorial system (martial and one party or one person system), constitutional royal system, and other political systems that are studied in depth by political thinkers. Despite the different political governments in the history, the most important event in recent decades is the conversion of these systems into democratic gathering, especially in the last and present centuries.

We aimed in this research to measure the political effects of information technology on nation and human life. People relate the political matters differently according to the political system being applied. As we mentioned in previous, most nations are governed by the democratic system. We would like to discover the playing role of this technology under these systems. Scientists are separated to three different groups in their view to this role. One group believe that the information technology increases the democracy of nations. Another group has completely different view to the information technology. They believe that the information technology works in contrary with democracy (Harper, 2003; Joint, 2005). The remainder of the scientists fall between the other two groups; they believe in neutral effect of information technology on democracy, people's democracy does not change even negatively or positively with this technology so it has no political effect on nations (Webster, 1999; Bentivegna, 2003).

In this study we will explore the information technology impact over the democratic practice of the people of Kuwait. This study is divided into the following parts: First, the historical background discusses the phases of the historical development of the political term *state* and how it arrives to its present form based upon theories and writings of the most influential thinkers in the field; namely Thomas Hobbes, Jean-Jack Rousseau and John Locke. The second part introduces definitions and models of democracy and its major components in political philosophy. In the third part we discuss the ruling practice, the constitutional regime components and the relationship between the state and its citizens in the State of Kuwait. After this point we transfer the attention to study the effect of technology on the democratic system in Kuwait. In the last part we step up to the research by a summarization beside the recommendations and the future researches.

## HISTORICAL BACKGROUND

By reviewing the political history, we notice that most political philosophers wonder how the life could be if people were living in the absence of government. Certainly, this discussion would help us reviewing the reasons behind law obedience and why people's choice set on identifying certain person or group to entertain this obeying right. This intellectual discussion further expands to a more in depth questionings for the obedience limits, as well the situations where the enforcement of such obedience by force becomes obligatory and necessary action. Moreover other issues are equally critical such as how the law can be enforced? Who have the authority to use violence? Can this authority be used against one particular group or all people? What are the circumstances that justify using violence?

To answer all these questions and more others, thinkers try to imagine how the people survive with no attendance of any governmental ruling. Man in the natural state adheres to the law of nature before he and other people form agglomerations and troops to preserve their rights as well as the rights of others. Although philosophers adopt this assumption in their theories, the well known French philosopher Jean-Jacques Rousseau argues against this hypothesis and claims no nature state occurred before the presence of government. On the contrary, Thomas Hobbes and after him John Locke assure this hypothesis. But far from discussing the truthfulness of this hypothesis, we think beginning the theoretical discussion from this state is a good tactics and intact debate. Its importance is due to the fact that the researchers assume the absence of government in order to justify its existence and appearance afterwards.

Another view comes from the English philosopher Thomas Hobbes (1588 - 1679) who sees that natural human being lives between two fires, fearing from the others surrounding him and the fire of greediness, mastery and domination (Wolf, 1996). Hobbes explains this idea in his famous book *Leviathan* in which he said that human being fears from others because everyone has in his deep-rooted the habit of greediness, which makes him aggressive to own the possessions of others. From another point of view, since man is self-centered and he is always prefers himself over others, he, therefore, loves and feels excited to be superior in all fields. For this reason he is always working hard to reach his main targets; richness and subjugation of the others. Hobbes adds that people are equals in strength, this equality means that killing and attacking others is an available choice for everybody including the weak. For this reason, all people live in unsafe continual fear and anxiety because of the possibility of being killed by others. Perhaps we may wonder the reasons behind intimidating each other? Hobbes responds that non stability and war in natural state is an affirmative because of the scarcity of goods. As a result, people always looking to what others have and waiting for an opportunity to take it from them. Hobbes concludes in his analysis that the state of nature with its assumptions (i.e. equality, scarcity of goods and uncertainty) is unstable as if it is a state of war.

According to the above analysis, Hobbes sees that moral values have no application in the State of Nature. There were no moral notions in societies before the existing of governments, no justice or altruism. Human being exchanged it for perfect freedom with aggression and murder either for self-defense or taking over others possessions. Under these circumstances, Hobbes arrives at a definitive conclusion by obtaining the necessity of the presence of a central government who granted the right of ruling people to protect them against any abuses or aggressions.

The other English philosopher John Locke (1632-1704) explains his theories of governance in his book Second Treatise. He agrees with Hobbes in his hypothesis of the perfect freedom in the natural state, but his analysis differs by the assumptions he uses. Beside what natural human being enjoys of perfect freedom, equality was found in societies before the existing of governments and there were also natural law that was applied by all people. Unlike Hobbes, Locke begins his analysis by his view to the moral root in natural human being in their dealings with others. Locke realizes that human being is naturally good and peaceful. Law of nature punishes offenders who harm another's life, property, or liberty. According to Locke, law of nature forbids people to use their power in harming others except in case of self-defense, in which situations, they can use their power. From this point Locke differs from Hobbes in his argument of freedom. For Locke, this perfect freedom in the state of nature is just for preserving the law of nature. Accordingly, we can see that Hobbes and Locke have significantly different views of the perfect freedom in law of nature in which people can do every thing to seek their needs through the moral values. For this reason, he believes that the law of nature morally justifies the use of penal code to prevent all attackers from harming others. The most significant different between the arguments of these two philosophers is related to the issues of abundance and scarcity of nature. Locke believes in the abundance of resources in nature, so people are not ambitious to get hold of other's possessions.

Hence, the state of nature is able to secure the peace of people in regard of two points: first, abundance of resources in nature, and second, the moral usefulness from punishing and fearing aggressors. Then, where did Locke stand from the constitutional government? Locke agrees with Hobbes in his view of the necessity of the government, though they vary on their justifications. In the times that Hobbes believes in an instability of the State of Nature (based upon the assumptions of scarcity and equality), Locke sees that forming the constitutional state is very important since it would resolve social related problems and apply justice. It is correct that people desire justice and use penal code for deterrence of aggressors, but Locke sees that the individuals are unable to have one explanation among all of them for the limits definitions of aggression, the suitable tools to enforce punishments, the manner in which the power is used, and how to diagnose criminals. From this point, the state is morally capable to apply justly the law of nature.

The French philosopher Jean-Jacques Rousseau (1712 - 1778) begins his theory of the state differently than Hobbes's. While he is certain that man of nature state is always looking for self-preservation and protecting his property from stealth, Rousseau believes that human beings are peaceful in their nature. Contradictory to Hobbes, Rousseau states that natural human being is peaceful by his nature, so he avoids violence and harming others, not because people are equal and enjoy the same circumstances and body strength as Hobbes argues, not also because there is plenty of resources in nature and fearing from natural penal code as Locke says, but instead, because man in his natural humanity trusts and feels the lust for peace, calm and compassion. He argues that violence or aggression is exceptional condition to human's peaceful nature. For Rousseau, people are always sympathetic to others, but they become abnormal by their suffering.

Rousseau believes that humanity cannot tolerate state of nature because God created man, the intellectual creature, who understands life through organized society where the government plays the role of the governor and the dominant of the authority. Although the hypothesis which is being relied on in the necessity of the government is in agreement with Hobbes in the scarcity of natural resources, he disagrees with him in his justification. Hobbes argues that scarcity and conflict of interests are the main reasons of problems and fights between people with one another causing wars. From another point of view, Rousseau says that scarcity is the major concern of humanity. So through his creative ability, man invented tools and machines in order to get out what is in the earth from treasures and fortunes. He noticed that the human being had begun developing machines to improve his ability in dealing with nature and to keep away the ghost of hunger away as possible specially with the continuous population growth. However, the results that he gets from using technology in repairing lands and getting out its fortunes made him live in welfare. After the fear of hunger, he became frightened from losing this welfare that were in excess of his needs, at this time human soul began to be infected by some negative values, such as private property, paupers, envy and the lack of the equality. Based on this debate, Rousseau arrives at the importance of establishing governments to enforce justice, where it aims to constitute the citizen's rights and equality to be legalized and entertained by the population.

## DEMOCRACY DEFINITIONS COMPONENTS

The Democratic system is one practice of ruling beside many other political systems; dictatorial, constitutional monarchist, and communist. We choose the democratic system to be the topic of this research for two reasons: The first is its worldwide acceptance nowadays, and second is due to the clear attention given by the contemporary studies of the philosophers in the field. Those scholars merge clearly in their studies between democracy and technology claiming that technology (specially the internet) is biased and carries democratic values in itself, which makes the discussion regarding these two concepts indispensable (Clift, 2000; Hoffman, 2000; Gronlund, 2001; Xenakis and Macintosh, 2005).

Democracy has a simple widespread notion among people. Moreover, it also implies another complicated definition that contains many partial clauses shaping the external appearance of the democracy. It is well known that democracy is the role for the people by the people and the role of the government is limited only to keep up people's property and rights. However this definition is extremely simplified, which makes the political philosophers differ clearly on their positions of its meaning. The difference is reflected clearly in their efforts that concluded into various and non-homogenous definitions and subtypes of democracy. Walzer (Walzer, 1965) divides the democracy to two main types: thin or minimalist versus thick or maximalist democracy. Meanwhile the first type is controlled by only general norms and principles, the second type is controlled by history, culture and society's customs.

The well known economist Dr. Joseph Schumpeter defines democracy as the essential forming for the political decision taking for which people compete to win the acceptance from nation (Schumpeter, 1975). However, Adam Brozotseker defines thin democracy as the control transference from one political group to another opposite political group through an election system respecting people's well at voting (Shaun, 2004). While, the famous political philosopher David Held (Held, 1997; Hacker and Dijk, 2001) analyzed democracy in his book "Models of Democracy" and divided it to nine types, where each type has its own features.

## EDUCATION, KNOWLEDGE AND DEMOCRACY

State of knowledge and education of the society are the main bricks for democracy. The knowledge level of the people considered the most important issue that occupied enormous attention from scholars and scientist of the philosophy of politics field. The old Greek philosophers; Plato, Aristotle and Socrates, whom were the first among scholars to study democracy, pointed out the importance of education development in democratic society, and for that reason they all agreed to limit election and voting to specific social classes in the ancient Greek society. This is also was adopted by the leaders of European Enlightment era, such as Thomas Hobbes, Jean-Jacques Rousseau and John Locke (Rosen, 2004; Reeve, 2004; Burns, 2004; Wolf, 1996; Rabeeh, 1994), John Dewey (Dewey, 1997a; Dewey, 1997b) and others, which was obvious in their writing and ideas.

Despite the numerous opinions pertaining democracy at past and present times, the present philosophers gave special attention to education when they studied this political concept (Watson and Mundy, 2001; Willinsky, 2002). An exceptional effort is made by the famous philosopher John Dewey who was known in linking democracy with education for society development and progress. Not only Dewey treated in his studies the vital and crucial linkage between education and democracy, but also he insisted in the importance of people's education to build up a democratic society (Dewey, 1997a; Dewey, 1997b). Willinsky expanded in his discussion the importance of education for society in present times. Present societies differ from the past one in that they are enjoying luxurious information technology, which was revolutionized through Internet. All of these reasons reduce the cost of education to a very low levels, which were considered a high standard through human history. The internet provides people with: low cost, short time to reach information, mass communication, recognizing variety of political opinions, and direct public-politicians debate. For these reasons, Willinsky insisted and magnified the importance of education through more participation over studies and research in different human science and social fields (Willinsky, 2002).

## KUWAIT AND DEMOCRACY

In this section we raise the issue of democracy in Kuwait and question if it is considered a democratic state? We dealt with the theoretical definition of democracy in philosophical view and how this definition had been developed during history. Then, we switched to knowledge and education and their effect on democracy. In this section, we need to understand how Kuwait performing its role of educating its people. In state of trying to reach the conclusion that Kuwait is a democratic state or not, we are urged to conclude whether if Kuwait awards education a special importance? Also, we try to answer other wondering questions: such as, is education in Kuwait suitable? What are other available resources for knowledge in Kuwait? And are Kuwaiti's lifes remarkably affected by technology?

Kuwait society has various information resources. One of these resources is education, which is supplied freely to the people. Education in Kuwait has been started in 1936/1937 with 600 students. By the late 60's of the last century, the number of students had reached about 80000 including male and female students. In year 1998/1999, there were about half million students. The education has been developed and the number of students is been increased due to establishments of special old-aged and the illiterates schools, which are established because of the new rules.

Beside the free education and the rule of diminishing illiterates, there are other different information resources for the people of Kuwait. The public libraries are a very important information resource. The first public library was established in 1924 by a group of citizens and they managed the library until year 1937, in which public libraries were attached to the ministry of education that was called the *Information Bureau* at that time. Kuwait gave a very special consideration to educate its citizens and that was reflected clearly when the royal act was issued on 17/7/1973 to establish the national council for culture, arts and literatures. The main goals of the council were as follow: 1) look to culture, 2) arts and literatures, and 3) maintain and develop intellectual and literal production. These goals are obtained through promoting different studies in various science fields, official adaptation of lexicons and dictionaries, and lastly, arranging competitions and passing prizes to the best literal and intellectual work.

Furthermore, there are other information resources available for the citizens of Kuwait that occupy significant role in educating people. Media and the press such as journals and magazines are considered the most attractive sources of information for the Kuwaiti population. Yet, the internet is the best source for information and knowledge. Kuwait is one of the premiers that use the internet especially among the Persian Gulf countries (Burkhart and Goodman, 1998). Even though, internet at its first years in Kuwait was limited to the academia and high income people, the usage of internet covers all area in Kuwait nowadays.

## CONCLUSION AND RECOMMENDATIONS

We discussed the theoretical definition of democracy in this research and we knew that democracy is debatable; it is not an issue which is agreed by all people but it is a philosophic political definition differs among different time, place and societies. Also we talked about many terms and the main components of democracy, and what we found is that there is a noticeable difference between the philosophers' definitions of democracy. Although there were varieties of contextualization of the term, all of them agree on education as a basic building block of democracy. Finally we discussed the real practice of democracy at the State of Kuwait. We found that Kuwait developed rules and regulations to achieve progress in educating society, which means that Kuwait democracy is reflected clearly at the education sector. However, we should be aware of the fact that this fact doesn't claim, accordingly, that Kuwait is a democratic country in all the aspects. For this reason, this research could be useful to start other new waves of researches in the future to discover new issues for democracy and how technology could affect it or get affected by it? Then, we could study Kuwait position and measure if Kuwait has remarkable achievement in the democratic scale.

## REFRENCES